\documentclass[10pt,twocolumn,letterpaper]{article}

\usepackage[pagenumbers]{cvpr} 

\usepackage{graphicx}
\usepackage{amsmath}
\usepackage{amssymb}
\usepackage{booktabs}

\usepackage[pagebackref,breaklinks,colorlinks]{hyperref}

\usepackage[capitalize]{cleveref}
\crefname{section}{Sec.}{Secs.}
\Crefname{section}{Section}{Sections}
\Crefname{table}{Table}{Tables}
\crefname{table}{Tab.}{Tabs.}

\def\cvprPaperID{*****} 
\def\confName{CVPR}
\def\confYear{2022}

\begin{document}

\title{Reference-Free Image Quality Metric for Degradation and Reconstruction Artifacts}

\author{Han Cui, Alfredo De Goyeneche, Efrat Shimron, Boyuan Ma, Michael Lustig\\
University of California, Berkeley\\
{\tt\small \{louiscuihan2018, asdegoyeneche, efrat.s, boyuan.ma, mikilustig\}@berkeley.edu}
}
\maketitle

\begin{abstract}
Image Quality Assessment (IQA) is essential in various Computer Vision tasks such as image deblurring and super-resolution. However, most IQA methods require reference images, which are not always available. While there are some reference-free IQA metrics, they have limitations in simulating human perception and discerning subtle image quality variations. We hypothesize that the JPEG quality factor is representatives of image quality measurement, and a well-trained neural network can learn to accurately evaluate image quality without requiring a clean reference, as it can recognize image degradation artifacts based on prior knowledge. Thus, we developed a reference-free quality evaluation network, dubbed "Quality Factor (QF) Predictor", which does not require any reference. Our QF Predictor is a lightweight, fully convolutional network comprising seven layers. The model is trained in a self-supervised manner: it receives JPEG compressed image patch with a random QF as input, is trained to accurately predict the corresponding QF. We demonstrate the versatility of the model by applying it to various tasks. First, our QF Predictor can generalize to measure the severity of various image artifacts, such as Gaussian Blur and Gaussian noise. Second, we show that the QF Predictor can be trained to predict the undersampling rate of images reconstructed from Magnetic Resonance Imaging (MRI) data.
\end{abstract}

\section{Introduction}
In recent decades, Computer Vision research has shown remarkable progress. A large body of work has demonstrated the powerful capabilities of Deep Learning (DL), especially Convolutional Neural Networks (CNNs), for various computer vision tasks. Studies demonstrated that neural networks could recognize both local and global features in images and understand the semantic relationship between objects and the background \cite{Authors25}, thus exhibiting strong performance in challenging Computer Vision tasks such as object detection\cite{Authors24} and instance segmentation\cite{Authors26}. 

\begin{figure}[t]
  \centering
   \includegraphics[width=1.0\linewidth]{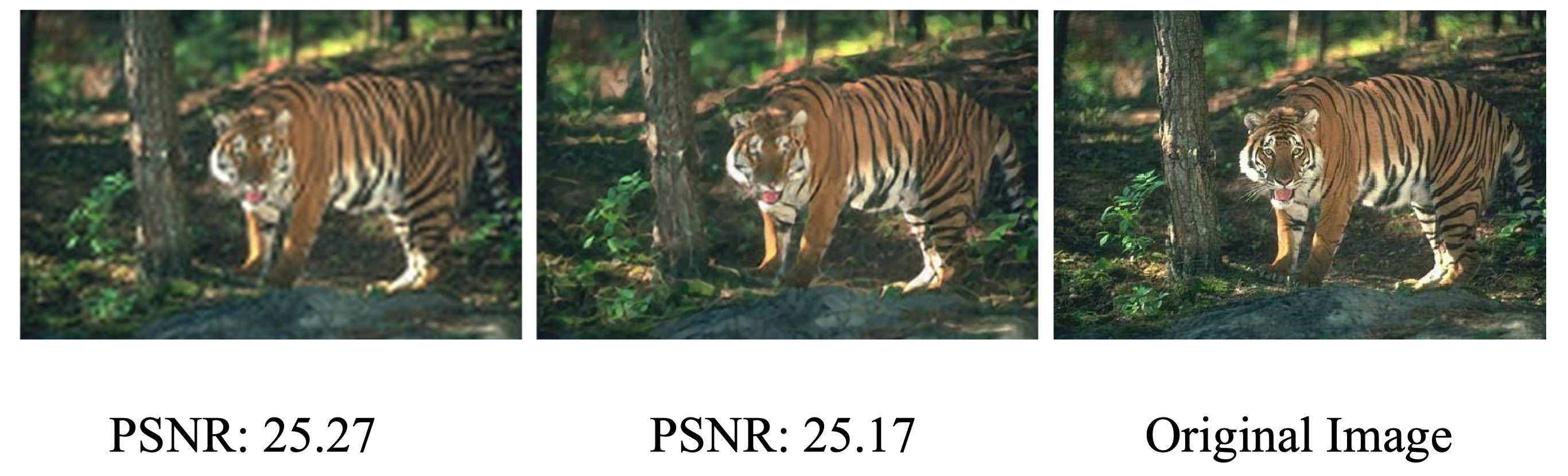}

   \caption{Examples that illustrate the discrepancy between PSNR values and human perception, adopted from\cite{Authors01}. Given the original image for reference, the left image has a higher metric score but a lower perceptual quality. On the contrary, the middle image has a lower metric score but a higher perceptual quality.}
   \label{fig:fig1}
\end{figure}

The development of DL models requires Image Quality Assessment (IQA) metrics,  as these metrics play a crucial role in the training process of the models. Two examples for such metrics are the Mean Square Error (MSE) and Structural Similarity Index (SSIM)\cite{Authors17}. These pixel-wise metrics are widely used in many applications\cite{Authors27, Authors29, Authors30}. However, their use  is limited in two aspects. First, computing such metrics requires a ground-truth image for comparison, but in many cases, such ground-truth data are not available. For example, in Magnetic Resonance Imaging (MRI), such ground truth images are difficult to obtain due to long scan time. Secondly, such metrics sometimes do not correlate well with human perception. For example, the Peak-Signal-to-Noise-Ration (PSNR) metric is typically used to measure image quality, where a higher PSNR score indicates better quality. However, in the example shown in Figure \ref{fig:fig1}, the PSNR metric assigns a higher score to the left image compared with the middle image. In contrast, human perception can immediately discern that the middle image is sharper and of better quality.

To address these limitations, we explore the development of a reference-free DL-based IQA model. The major contributions of this work are: 

\begin{enumerate}
  \item We develop a Deep Learning method for reference-free IQA, employing a self-supervised training approach with JPEG compression QF to avoid the need for a massive amount of manual data annotation.
  \item We design a network that learns to identify subtle JPEG compression artifacts, enabling it to differentiate nuances between fine-quality images and high-quality images.
  \item Our lightweight, CNN-based model can generalize to estimate the severity of other (non-JPEG) image degradation artifacts, including blurring, additive Gaussian noises, and imperfections in Magnetic Resonance (MR) images reconstructed from under-sampled data. 
\end{enumerate}

\begin{figure*}[t]
  \centering
  \includegraphics[width=1.0\linewidth]{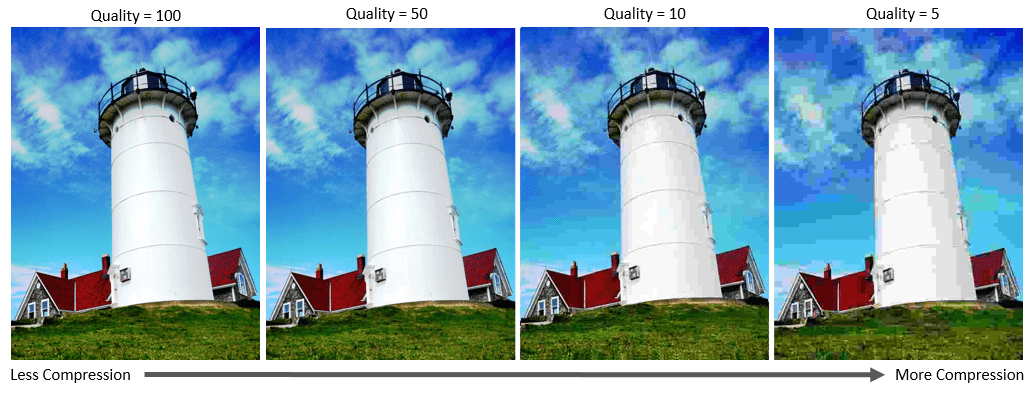}

  \caption{Showcase of JPEG Compression artifacts within different QFs.\cite{Authors09}}
  \label{fig:fig2}
\end{figure*}

\section{Related Work} \label{Relatedwork}

\textbf{Deep IQA.} Given the versatility of DL in various Computer Vision domains, researchers have already begun incorporating Deep Learning methodologies for automated IQA. These methods demonstrate excellent performance in quantifying the similarity between images\cite{Authors03, Authors04}, and modeling subjective human evaluations of image quality\cite{Authors05, Authors07}, particularly in the context of pairwise comparisons\cite{Authors06}. Most of these contributions focus on perceptual image quality. However, they can produce misleading results for slightly corrupted images. Also, many of these studies are based on very large neural networks, which perform optimally only when a substantial amount of manually annotated data is available.

\textbf{Self-Supervised Learning.} To reduce the need for large training datasets, researchers investigate methods for generating labels from the data itself, an approach known as Self-Supervised Learning. Many studies demonstrate that carefully crafted self-supervision training enables network to acquire high-level knowledge and features without human annotation. Examples include training networks to solve jigsaw puzzles\cite{Authors19}, to colorize grayscale images\cite{Authors18}, and to predict the angle of image rotations\cite{Authors08}. Specifically, networks trained for the latter task  show superior performance in downstream classification tasks compared to state-of-the-art algorithms, while also requiring significantly less training data. 

\textbf{JPEG-Compression.} JPEG compression attracted substantial research interest, as JPEG is one of the most widely used compression techniques. JPEG compression builds upon the Discrete Cosine Transformation (DCT) and uses the variable quality factor based on Quantization to control the compression ratio. Although effective for compression, JPEG also introduces some artifacts. As shown in Figure \ref{fig:fig2}, it generally produces a complex combination of blurring, blocking artifacts, and ringing effects\cite{Authors10}. Moreover, numerous studies, including both digital image processing techniques\cite{Authors22, Authors23} and DL-based approaches\cite{Authors20, Authors21}, have focused on the restoration of low-quality JPEG compressed images and the reduction of severe JPEG compression artifacts.

\textbf{Medical Image Restoration.} Medical images generally contain many detailed textures and require high image quality for diagnosis. Consequently, significant research efforts were dedicated to image restoration, focusing on the tasks of medical image denoising\cite{Authors31, Authors32} and reconstruction\cite{Authors33, Authors34}. Specifically, medical image reconstruction is the process of generating a high-quality and accurate image from a series of acquired measurements. 

A widely used architecture used for medical image restoration is the U-Net\cite{Authors28}, a convolutional neural network originally designed for image segmentation, which has been adapted for various biomedical restoration tasks due to its ability to capture both local and global features in the input images. A different, more advanced architecture is physics-guided unrolled neural networks. Such networks unroll an optimization algorithm and iterate between data consistency blocks and denoising blocks. One notable example of an unrolled reconstruction network is Model-based Deep Learning (MoDL)\cite{Authors11}, which was developed for MRI reconstruction. MoDL has shown a great capability to reconstruct high-quality images from under-sampled k-space data.

\section{Approach}

\subsection{Methodology Overview}

The goal of this work is to develop a Deep Learning method capable of evaluating IQA metrics in a reference-free manner. Our model is trained using a self-supervised approach as follows: we apply the JPEG compression to each input image with a random target Quality Factor (QF), which produces the compressed image. The network takes in the compressed input image, and outputs the predicted QF, which should be as close as possible to the target QF. This forms a simple optimization objective: 

\begin{equation}
\label{eq:1}
\begin{aligned}
\min_{\theta} \quad & \frac{1}{N}\sum_{i=1}^{N}{loss(y_{i}, y'_{i})}\\
\textrm{s.t.} \quad & X' = J(X, y)\\
  & y'_{i} = F(X'|\theta)\\
\end{aligned}
\end{equation}

In this formulation, $J(.)$ denotes the JPEG Compression process, $X$ is the original image, $y$ is the target QF, $X'$ is the compressed input image, $F(X'|\theta)$ represents the network output given $X'$ as input, $\theta$ are the network parameters, and $y'$ is the predicted QF. Our general formulation enables using different loss functions,  such as the Cross-Entropy Loss for a classification problem or the Mean Square Error (MSE) for a regression problem.

\begin{figure}[t]
  \centering
   \includegraphics[width=1.0\linewidth]{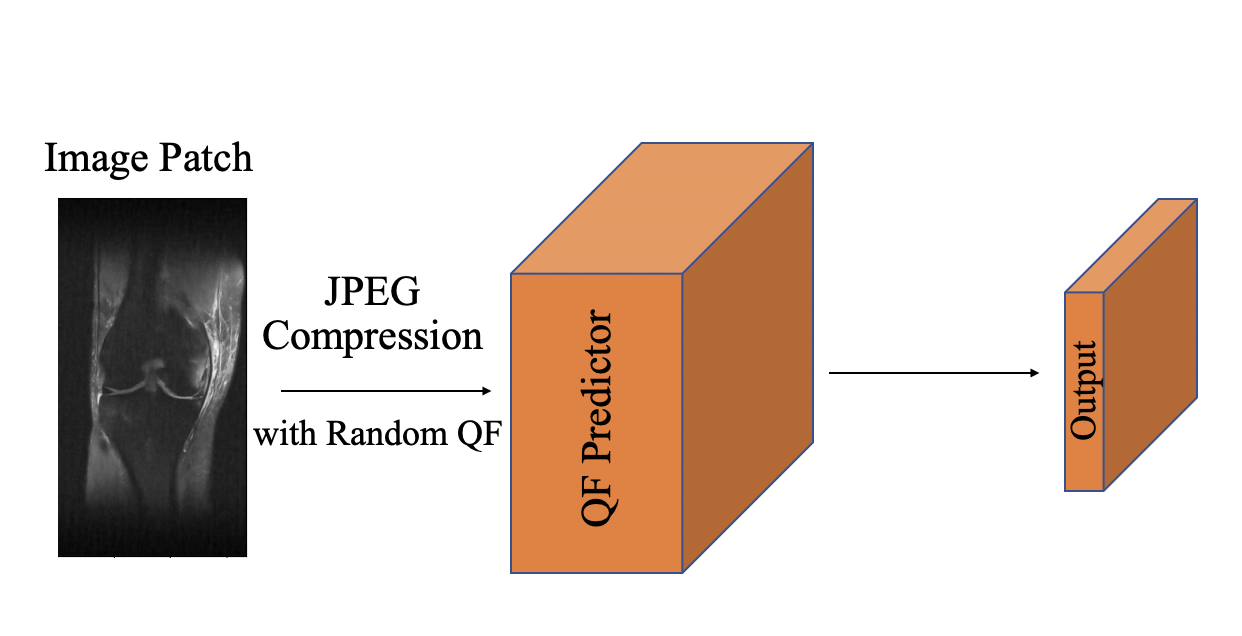}

   \caption{Training pipeline for QF Predictor}
   \label{fig:fig3}
\end{figure}

In Figure \ref{fig:fig3}, we show a simplified version of our training pipeline. The input image can be either an RGB image or a gray-scale MR image. In order to train the network with one specific image shape but be able to run inference tasks with different image sizes, we built the network in a fully convolutional way. In our setup, the desired output is a map of QFs, with a size smaller than the input image. Each value in the QF map represents the QF prediction for a specific region in the original image. Specifically, each value of the output QF map has an approximate field of view (FOV) of 11 pixels in the input image.

\subsection{Network Architecture and JPEG QF Values}
\begin{figure*}[t]
  \centering
   \includegraphics[width=1.0\linewidth]{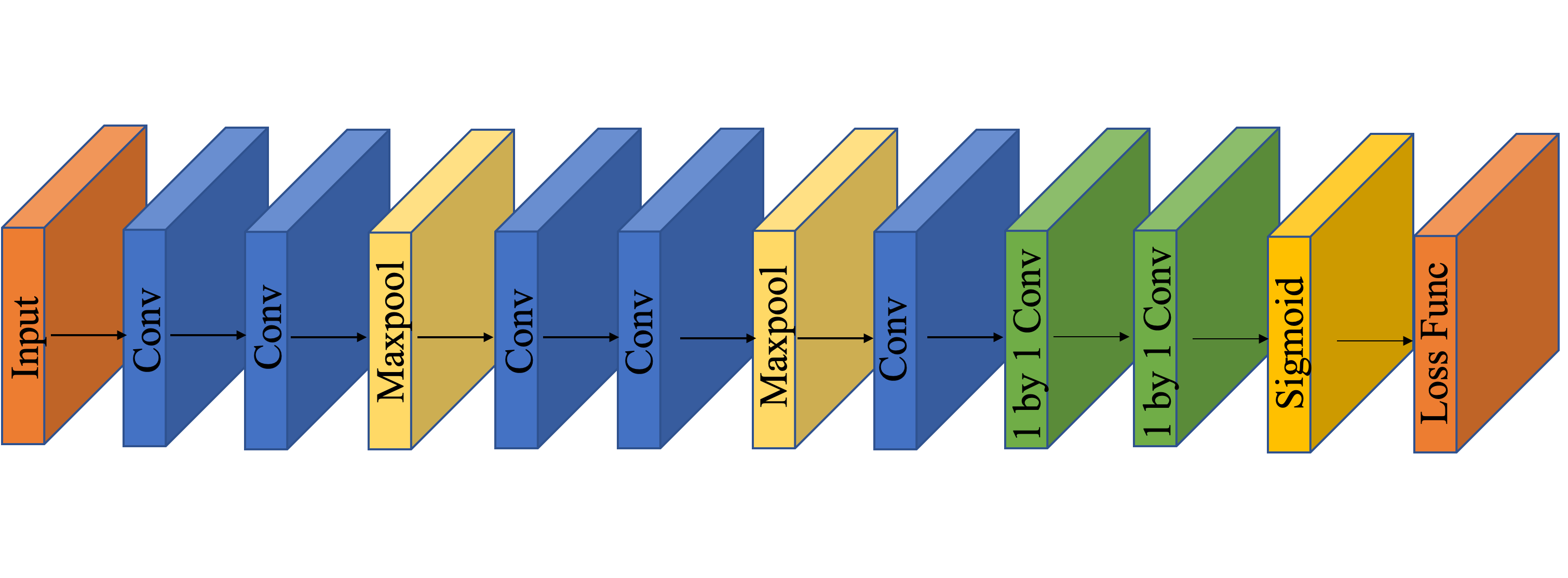}

   \caption{QF Predictor Architecture} 
   \label{fig:fig4}
\end{figure*}

In this project, we aim to develop a compact model. Therefore, we built a simple DNN with a small number of parameters, which has only seven convolution layers, as shown in Figure \ref{fig:fig4}. In the first several convolution layers, we incorporated two max-pooling layers to reduce the size of tensors and the computational complexity. In between every two layers, we included batch normalization and ReLU activation to introduce non-linearity. Among these seven layers, the last two layers utilize 1x1 convolutions to reduce the number of channels. 

The range of possible JPEG compression QF values is [0, 100], where 0 indicates the maximum compression level, and 100 indicates the minimum level. To make the optimization problem described in equation (\ref{eq:1}) compatible with general Machine Learning loss functions, which usually receive values in the range of 0 to 1, we re-scaled the JPEG QF range to [0.0, 1.0]. To avoid negative QF output, we introduced an extra Sigmoid activation level after the final convolution layer; this enforces the [0.0, 1.0] range. The network can automatically scale with the Sigmoid function, which does not harm the overall performance. 

\subsection{Datasets}
As mentioned earlier, we want the network to capture the perceived image quality from a human observer's perspective and find indistinct imperfections in high-quality images. For this purpose, we need high-quality datasets for training. Our interest lies in estimating the artifact level for both general RGB images and MR images reconstructed from under-sampled data. Thus, we trained two networks: an RGB QF Predictor and a grayscale QF Predictor.

The RGB QF Predictor was trained on the \textbf{\textit{Flickr1024}} \cite{Authors12},  a high-resolution dataset originally used for large-scale Stereo Super Resolution task. The grayscale QF Predictor was trained on \textbf{\textit{FastMRI}}\cite{Authors13}, a large database of raw MRI data. We utilized the knee dataset of \textit{FastMRI}, as it offers high-quality ground truth images. Additionally, we also run generalization experiments on \textbf{\textit{ImageNet}}\cite{Authors02}, a large-scale and widely-used dataset designed for training and benchmarking various machine learning and computer vision algorithms, and \textbf{\textit{LIVE}}\cite{Authors14}, a comprehensive IQA dataset consisting of a variety of distortion types, including JPEG and JPEG2000 compression, Gaussian blur, and additive white Gaussian noise.

Our work requires the network to identify small details in the images. Therefore, rather than feeding the whole image into the network, we randomly cropped small patches for training to ensure our network is sensitive to artifacts present at the patch size level. This can also be regarded as a data augmentation method to provide more randomness in the training procedure.

\begin{figure*}[t]
  \centering
  \includegraphics[width=1.0\linewidth]{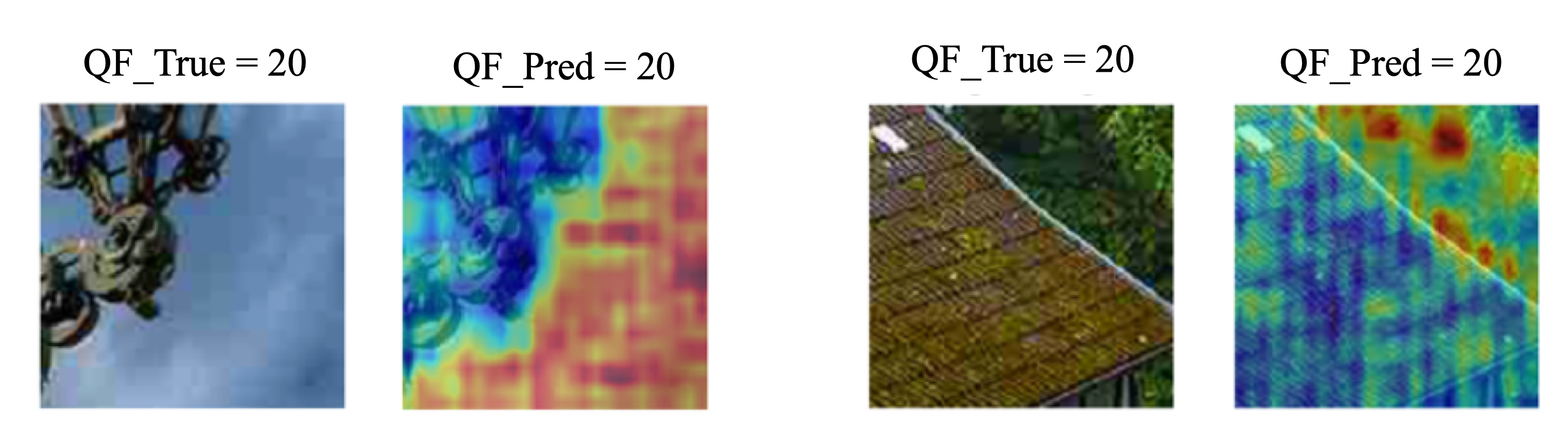}

  \caption{Examples of activation maps extracted from the last convolution layer. These results indicate that the network over-emphasizes the grid-like patterns in the corrupted images, which leads to biased results toward low QFs. }
  \label{fig:fig5}
\end{figure*}

\begin{figure}[t]
  \centering
   \includegraphics[width=1.0\linewidth]{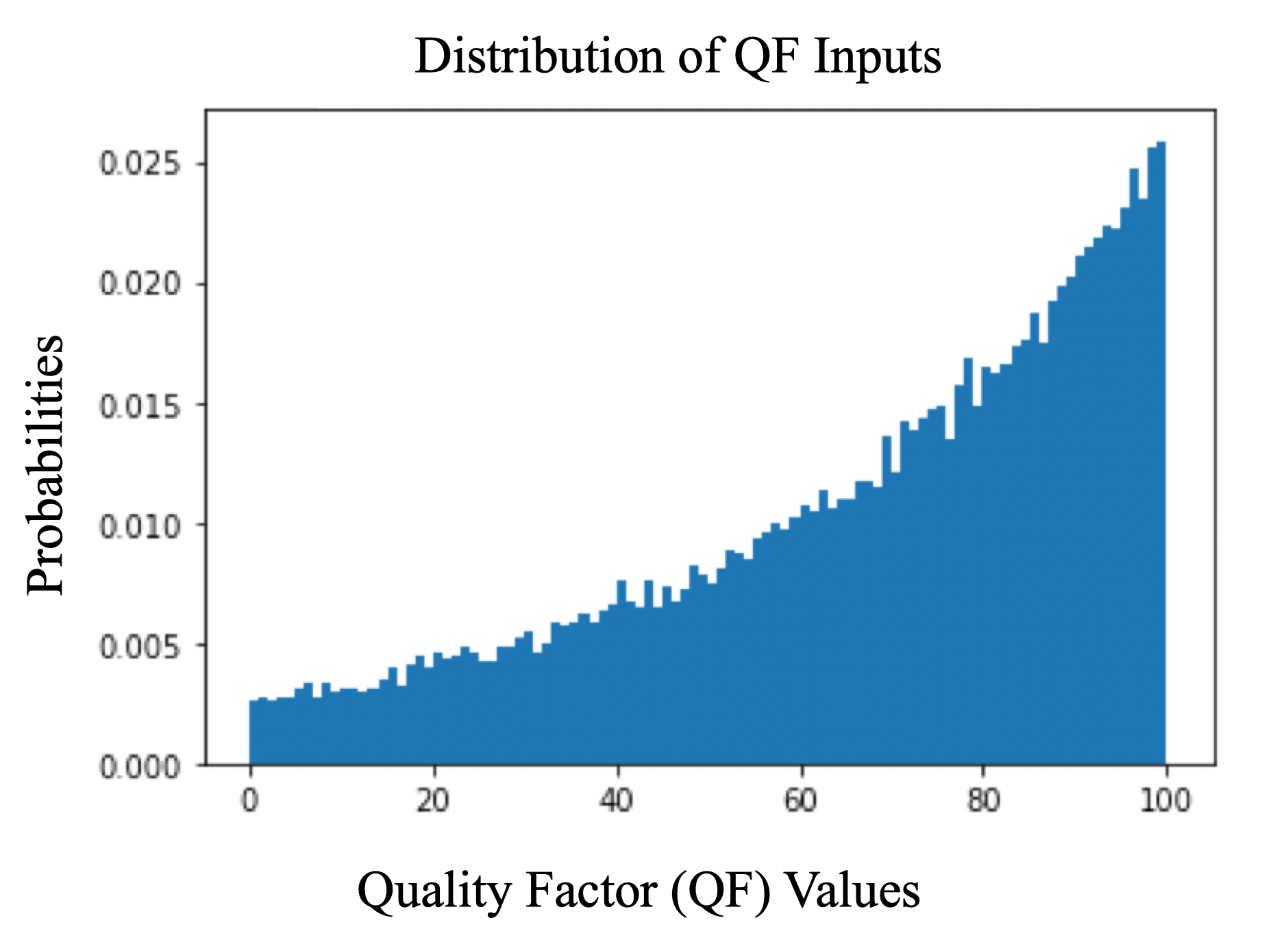}

   \caption{Weighted distribution for high-quality focusing training} 
   \label{fig:fig6}
\end{figure}

\section{Exploration of Training Approaches}

In this section, we describe the various approaches and engineering choices that were examined, along with an analysis of the rationale behind these decisions. In all the experiments described here, our RGB QF Predictor was tested on the validation data from \textbf{\textit{Flickr1024}}.

\subsection{Training with a weighted QF distribution}
 As noted earlier, we apply JPEG compression to image patches compressed with a randomly selected QF, and then train the network to predict the ground truth QF. In our first set of experiments, the random QFs were drawn from the range [0, 100] using a \textbf{random-uniform distribution}. However, we observed that this design resulted in a model that exhibited proficient performance for low QF values, in the range of [0, 40] and suboptimal performance for QFs in the range of [80, 100]. To examine the reasons, we plot the model's activation maps, which were extracted from the last convolution layer (Figure \ref{fig:fig5}). As indicated in the literature \cite{Authors10}, JPEG compression generally produces a complex combination of different artifacts. However, we observed that the blocking artifacts become dominant for low QFs (Figure \ref{fig:fig5}). Because these artifacts are trivial for the network to identify, that leads to \emph{biased results}, where the QF is predicted accurately mostly for the low QFs regime. To solve this problem, we replaced the uniform distribution with a \textbf{non-uniform} one, shown in Figure \ref{fig:fig6}. This distribution was generated by assigning probability based on the logarithmic value of QFs. It therefore assigns higher weights to QFs in the range of [80, 100]; as a result, during training, QF values were mostly drawn from this range. This forces the network to focus on learning rare artifacts that appear mostly in slightly-compressed images, while still being exposed to blocking artifacts (that appear for highly compressed images) as early-stage guidance. As a result, we achieve our goal of differentiating nuances between high-quality images and lossless images.

\subsection{Classification vs. Regression}

\begin{figure}[t]
    \begin{subfigure}[b]{0.45\textwidth}
         \centering
         \includegraphics[width=1.0\textwidth]{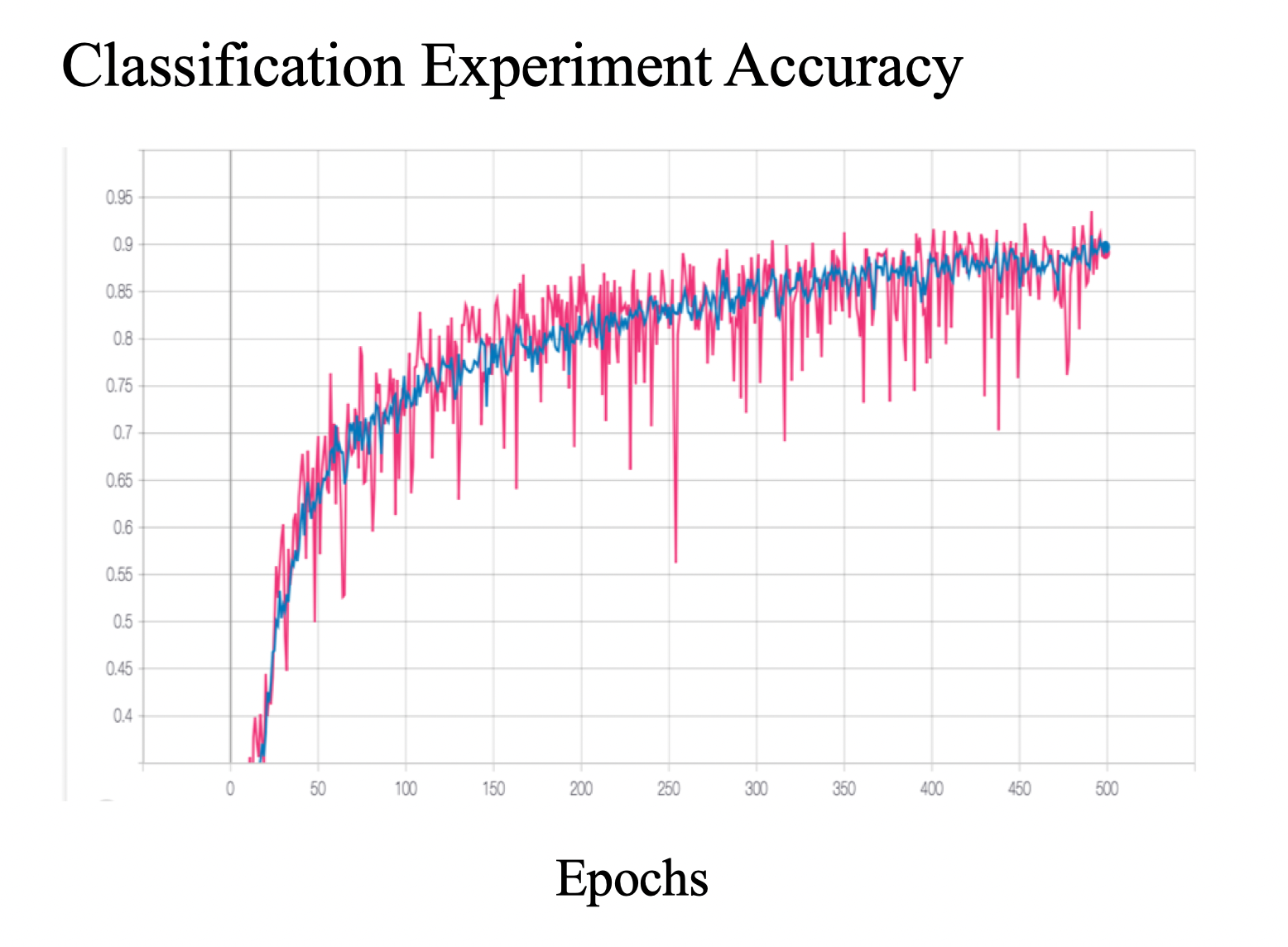}
         \caption{Curves for Cross-Entropy}
         \label{fig:7a}
    \end{subfigure}
    \begin{subfigure}[b]{0.45\textwidth}
         \centering
         \includegraphics[width=1.02\textwidth]{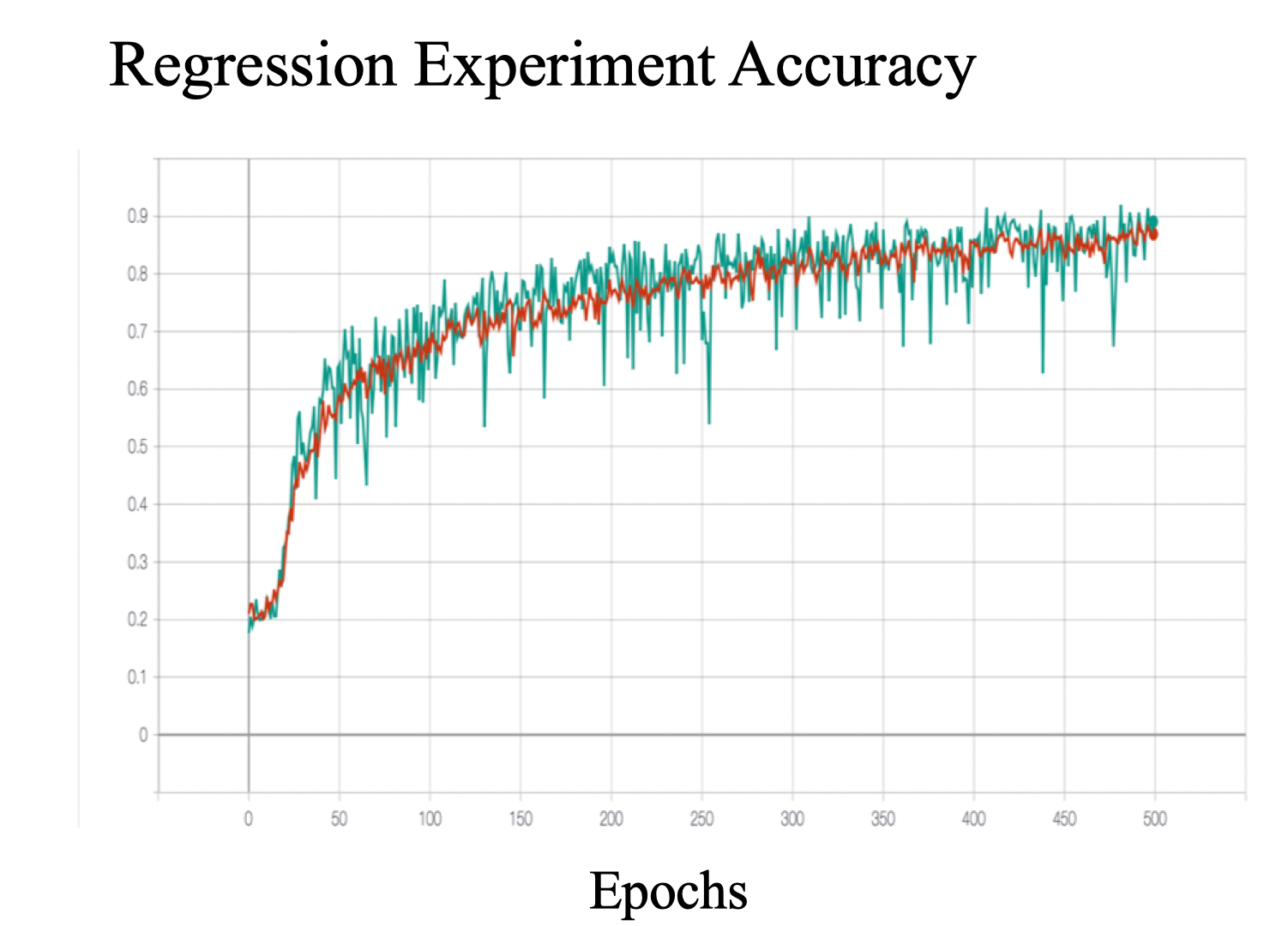}
         \caption{Curves for MSE}
         \label{fig:7b}
    \end{subfigure}
    \caption{Training and Validation curves for two loss choices.}
   \label{fig:fig7}
\end{figure}

\begin{figure}[t]
    \begin{subfigure}[b]{0.47\textwidth}
    \includegraphics[width=1.0\textwidth]{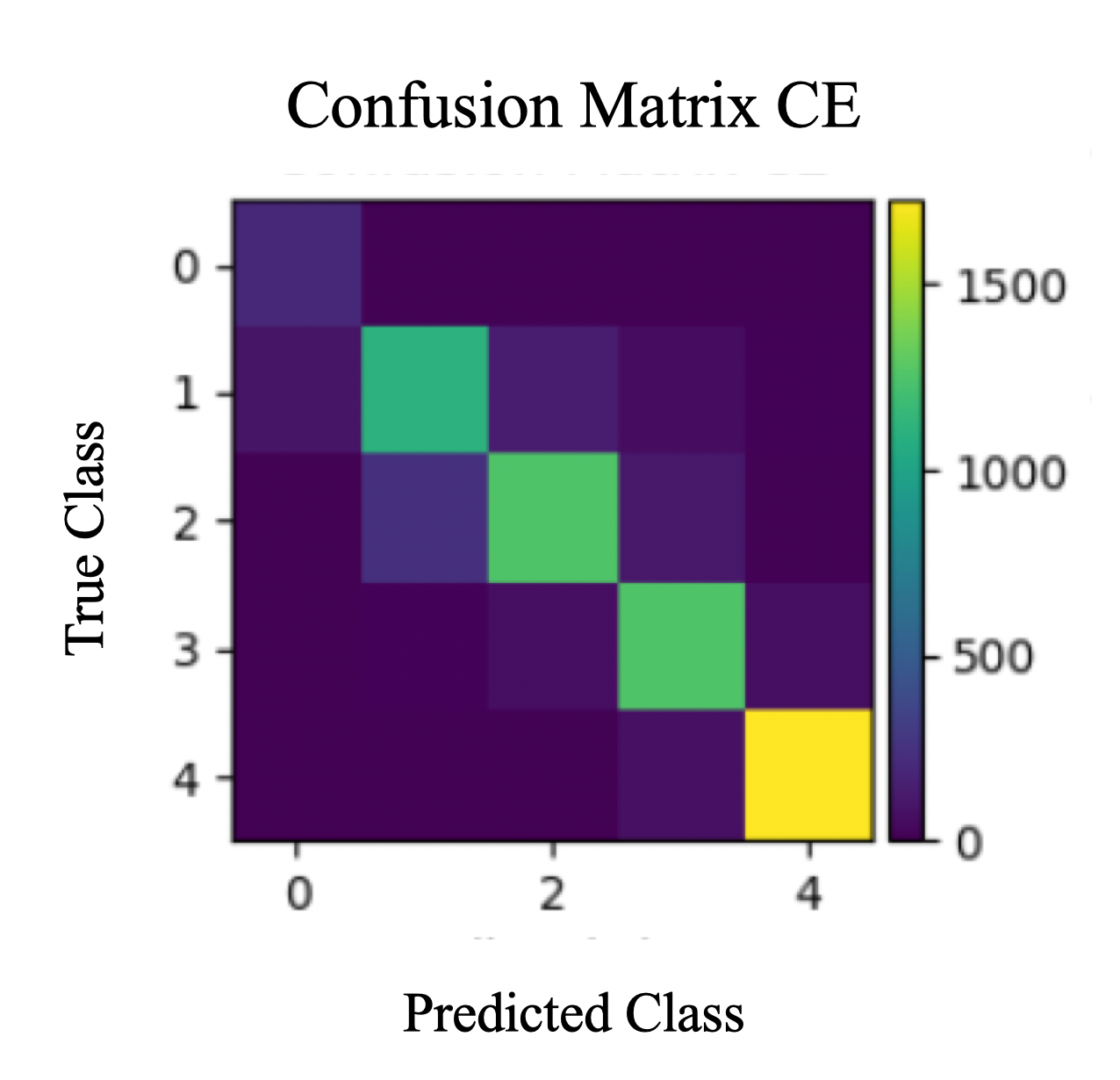}
    \caption{Confusion Matrix for Cross Entropy validation data} \label{fig:8a}
  \end{subfigure}
  \begin{subfigure}[b]{0.47\textwidth}
    \includegraphics[width=1.0\textwidth]{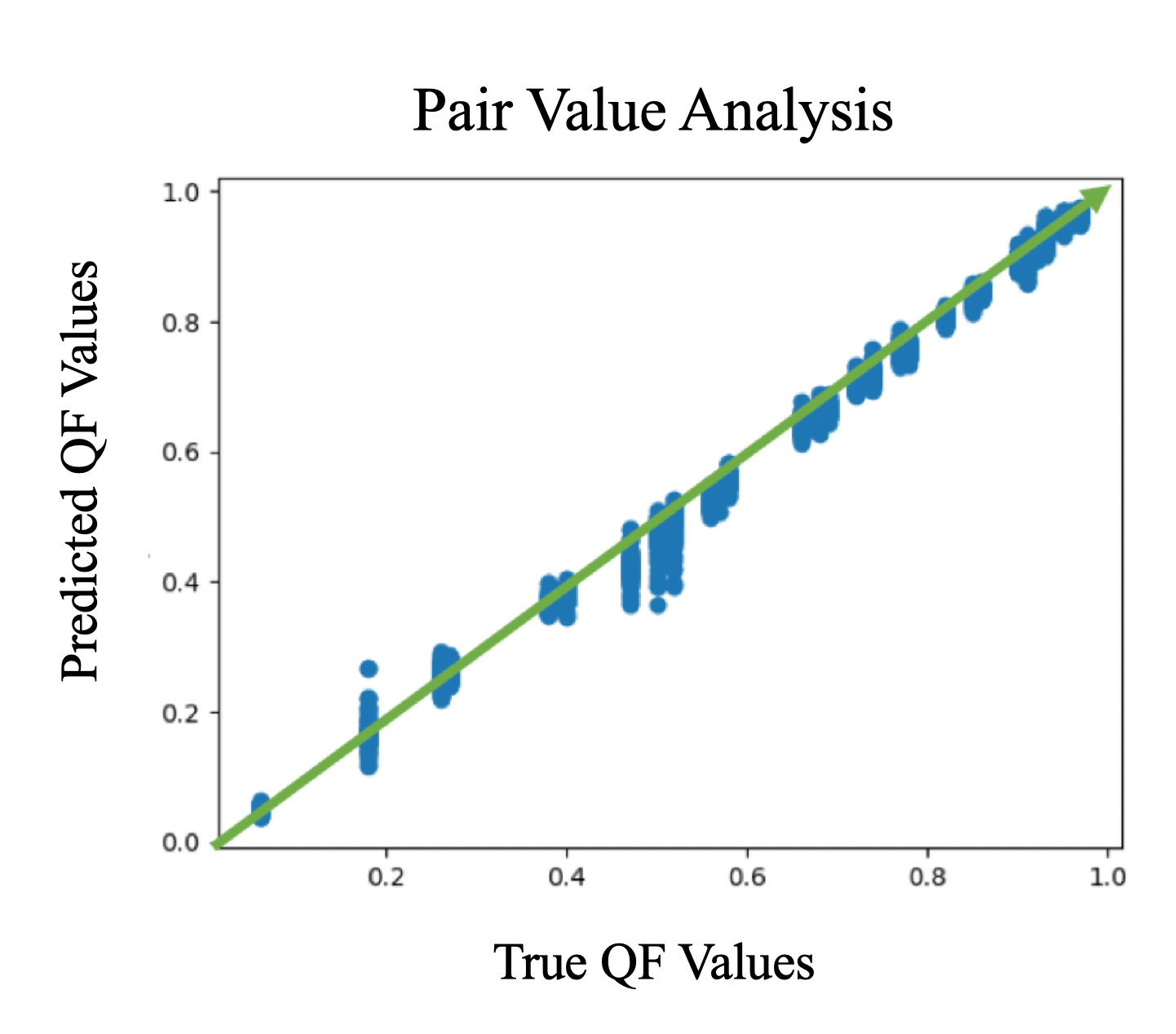}
    \caption{Pair Value Analysis for MSE validation data} \label{fig:8b}
  \end{subfigure}
    \caption{Extra Visualizations for two loss choices.}
   \label{fig:fig8}
\end{figure}

As mentioned earlier, the choice of the loss function in Equation (\ref{eq:1}) might be an important engineering choice in the project. As a first step, we formulated the problem as a classification task with the Cross-Entropy loss as the following: we only select QFs from [0, 20, 40, 60, 80], with more training examples focusing on the case where QF = 60 or 80. We can treat each QF as a class category and treat the optimization problem as a classification problem with five categories, where class 0 corresponds to QF 0, class 1 corresponds to QF 20, and so on. This setup forms a simple experiment that can help us understand the network's capacity and tweak hyperparameters, such as the size of the image patch and the learning rate. However, treating the problem as a classification task poses a potential issue: the network is not required to learn sophisticated knowledge to distinguish between examples with QF = 60 and those with QF = 80, which hinders its sensitivity to the subtle differences between fine-quality and high-quality images.

\begin{figure*}[t]
   \begin{subfigure}{0.5\linewidth}
    \includegraphics[width=0.98\linewidth]{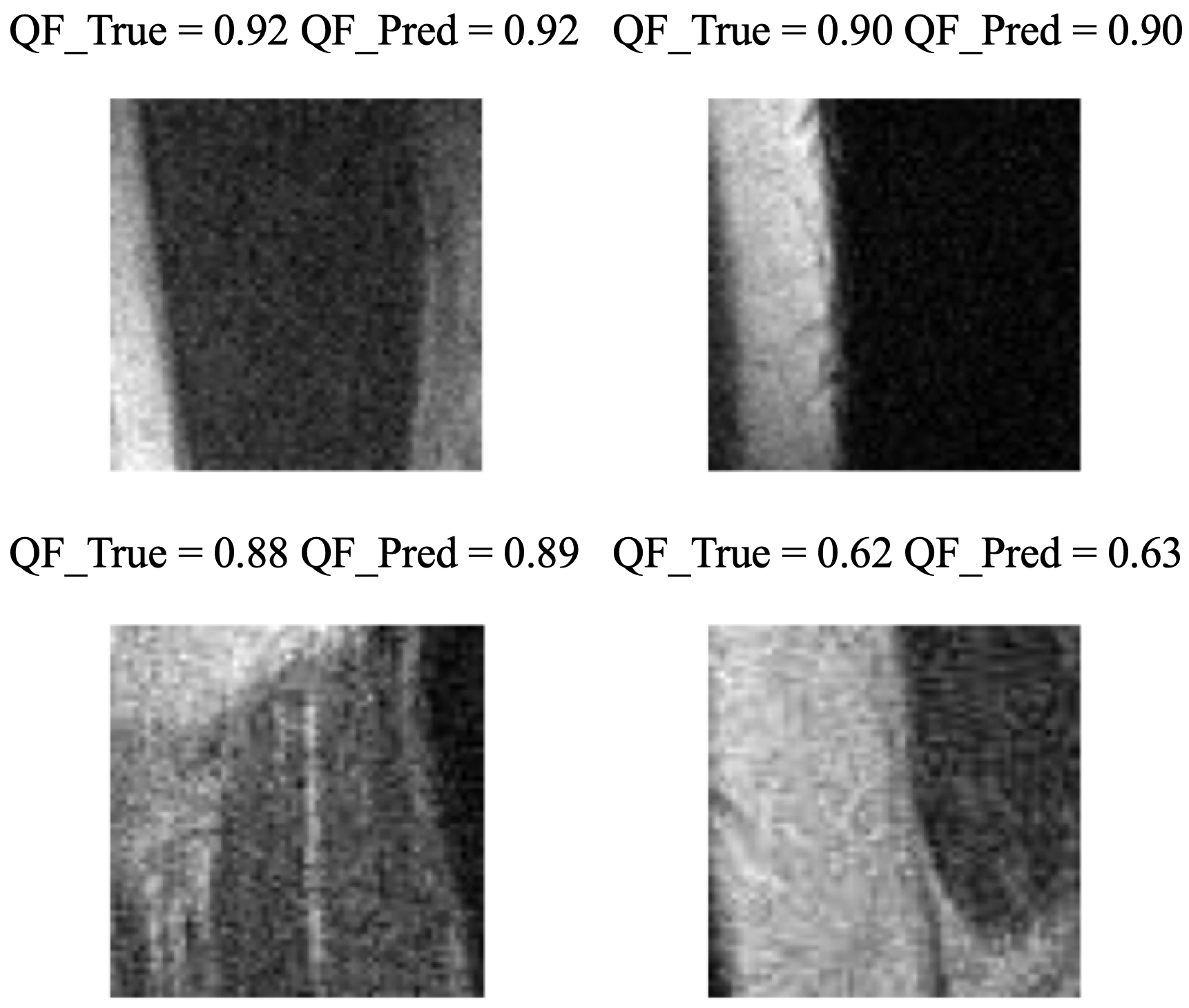}
    \caption{Grayscale examples} \label{fig:9a}
  \end{subfigure}%
  \begin{subfigure}{0.5\linewidth}
    \includegraphics[width=1.04\linewidth]{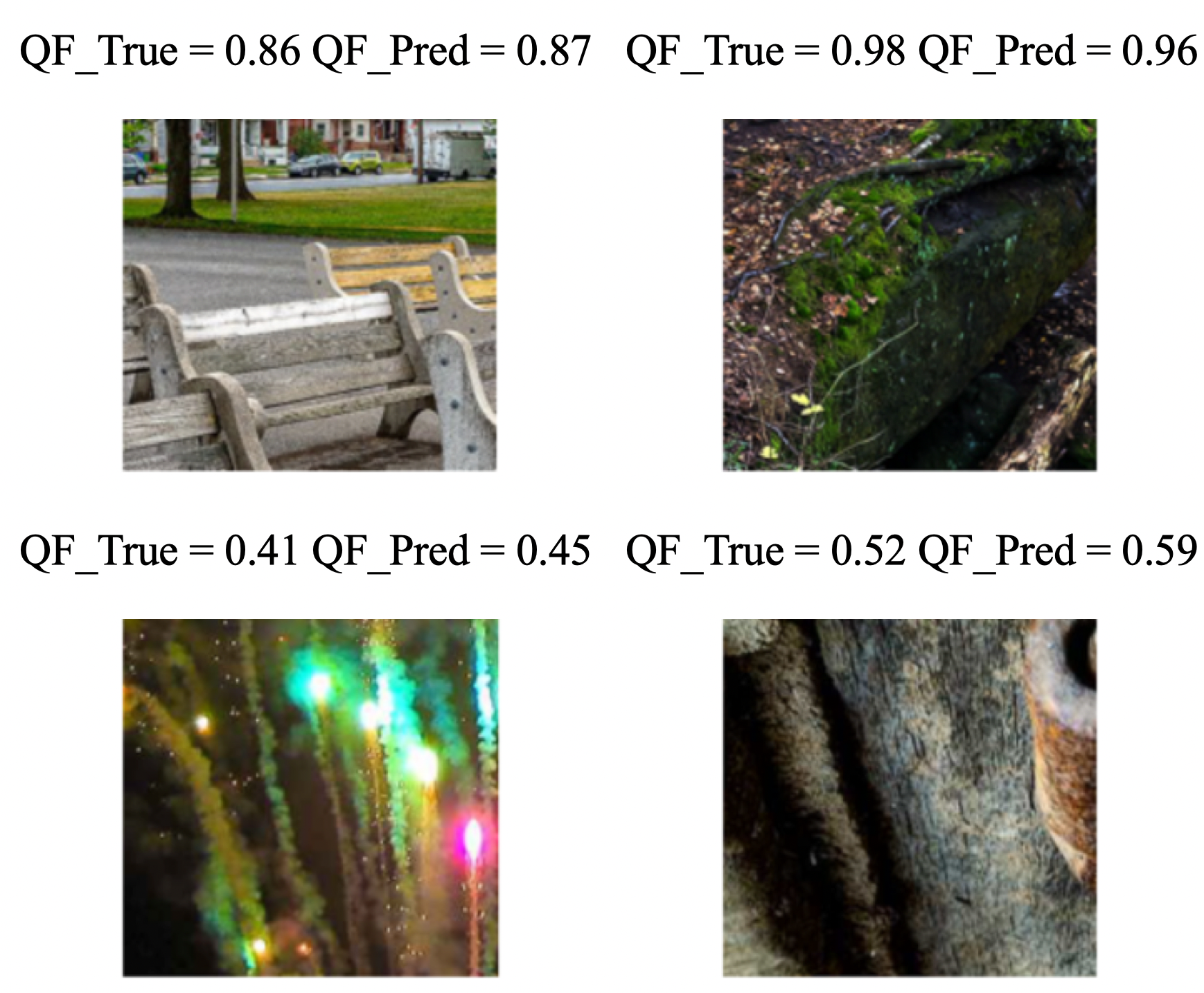}
    \caption{RGB examples} \label{fig:9b}
  \end{subfigure}%
    \caption{Validation Examples with ground truth and predicted QFs.}
   \label{fig:fig9}
\end{figure*}

Therefore, we decided to reformulate the problem. Instead of treating the QF prediction as a classification task, we decided to address it as a regression task. The regression task covers a broader range of QFs, introducing increased randomness into the training process and resulting in a more robust model. In other words, instead of predicting QF values into groups, we decided to train the network to predict the QF value on a continuous axis. As mentioned above, we rescaled QF values from [0, 100] to [0.0, 1.0]. In this experiment, we used the same architecture as before (Figure \ref{fig:fig4}), and trained the network  utilizing the Mean Square Error (MSE) loss, with the same training data from \textbf{\textit{Flickr1024}}.  

The training and validation curves for both the classification experiment (Cross-Entropy loss) and regression experiment (MSE loss) are shown in Figure \ref{fig:fig7}. Even though the validation curves oscillate more severely than the training curves in both scenarios, the overall training converges approximately after 500 epochs. Note that for the regression problem with the MSE loss, the accuracy in the graph is calculated as follows: the case is considered correct only if $|y_i-y_i'| <= 0.02$ where $y_i$ is the ground truth QF and $y_i$ is the predicted QF. Intuitively the regression problem is much more difficult than the classification problem, which explains the lower accuracy. 

For more straightforward visualizations, we include a confusion matrix for Cross-Entropy data in \ref{fig:8a}, and a pair value analysis graph for MSE data in \ref{fig:8b}. In both examples, the network  exhibits uncertainty about data with Quality Factors 40 to 60 due to the lack of training. However, the prediction for high-quality data, i.e. data that has QF of 80-100, exhibits a low error rate and low variance because of more training examples, which indicates the network performs accurately with a high confidence level on these high-quality data. In Figure \ref{fig:fig9}, we include some examples from the validation data evaluated by these two pre-trained models. These results show that the network has a stable performance and predicts close QFs to the ground truth for individual examples in both datasets.  

\section{Experiments} 

For the following inference experiments, we employ two pre-trained networks: RGB QF Predictor,  trained using the MSE loss and the \textbf{\textit{Flickr1024}} dataset, and the grayscale QF Predictor, trained using the same loss and the \textbf{\textit{FashMRI}} dataset.

\subsection{Generalization Test to other Artifacts}

\begin{figure}[t]
  \centering
   \includegraphics[width=1.05\linewidth]{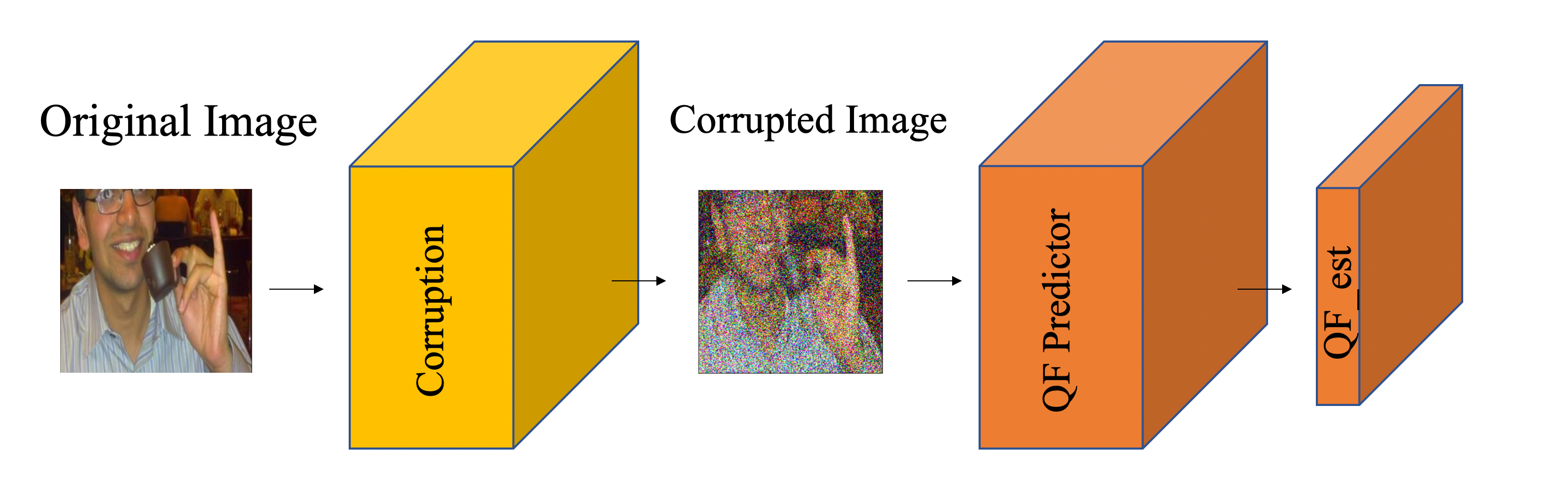}

   \caption{Inference pipeline for RGB QF Predictor}
   \label{fig:fig11}
\end{figure}

\begin{figure*}[t]
    \begin{subfigure}{0.5\textwidth}
    \includegraphics[width=1.0\textwidth]{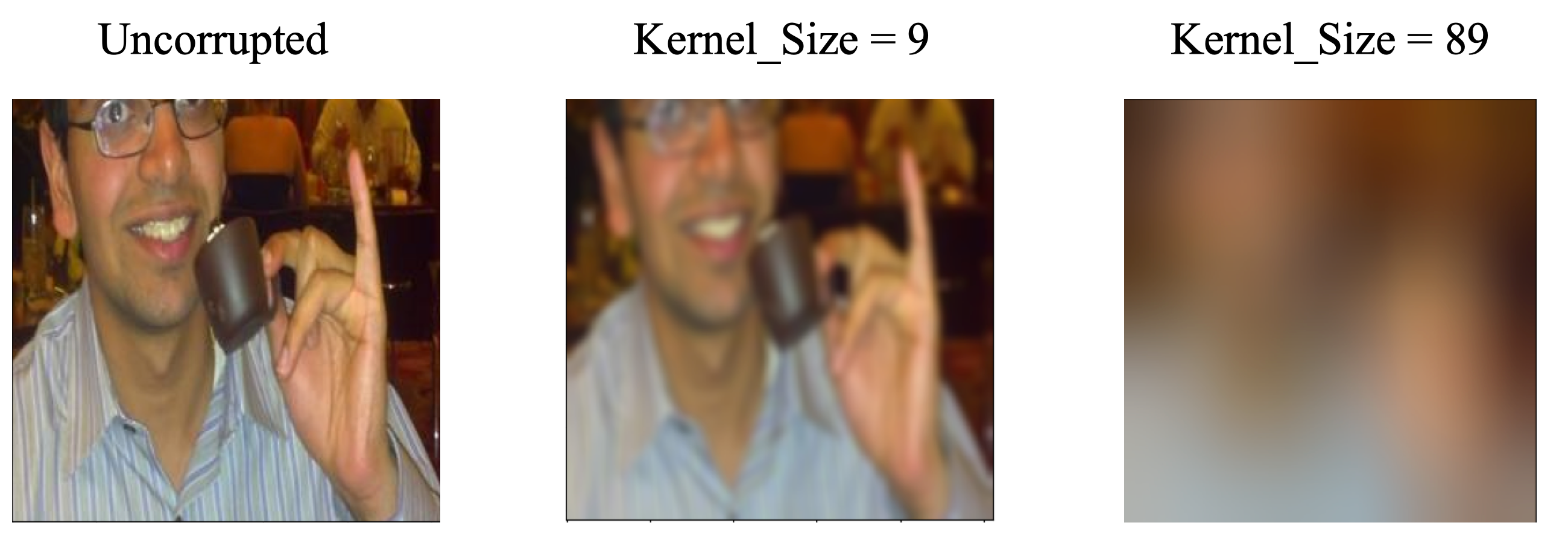}
    \caption{Blurring artifacts on LIVE \cite{Authors14} dataset} \label{fig:10a}
  \end{subfigure}
  \hspace*{\fill}
  \begin{subfigure}{0.5\textwidth}
    \includegraphics[width=1.0\linewidth]{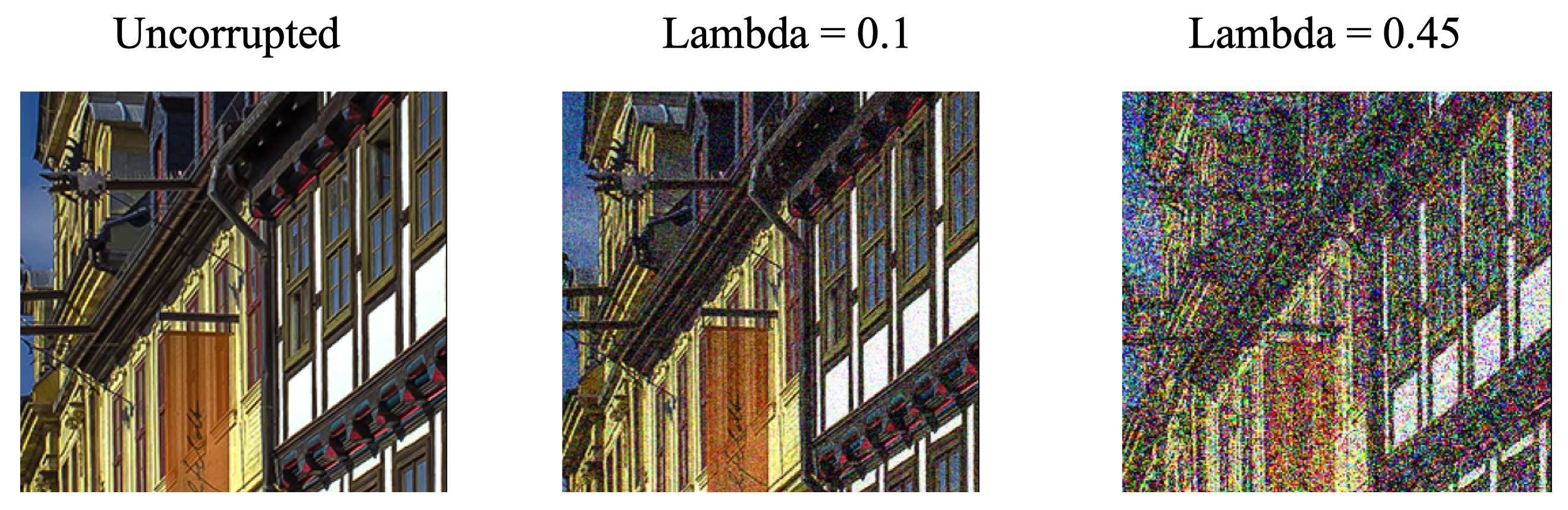}
    \caption{"Salt \& Pepper" artifacts on Flickr1024 \cite{Authors12} dataset} \label{fig:10b}
  \end{subfigure}
    \caption{Examples for blur and noise artifacts used in the generalization test.}
   \label{fig:fig10}
\end{figure*}

\begin{figure*}[t]
    \begin{subfigure}[b]{0.5\textwidth}
         \centering
         \includegraphics[width=1.0\textwidth]{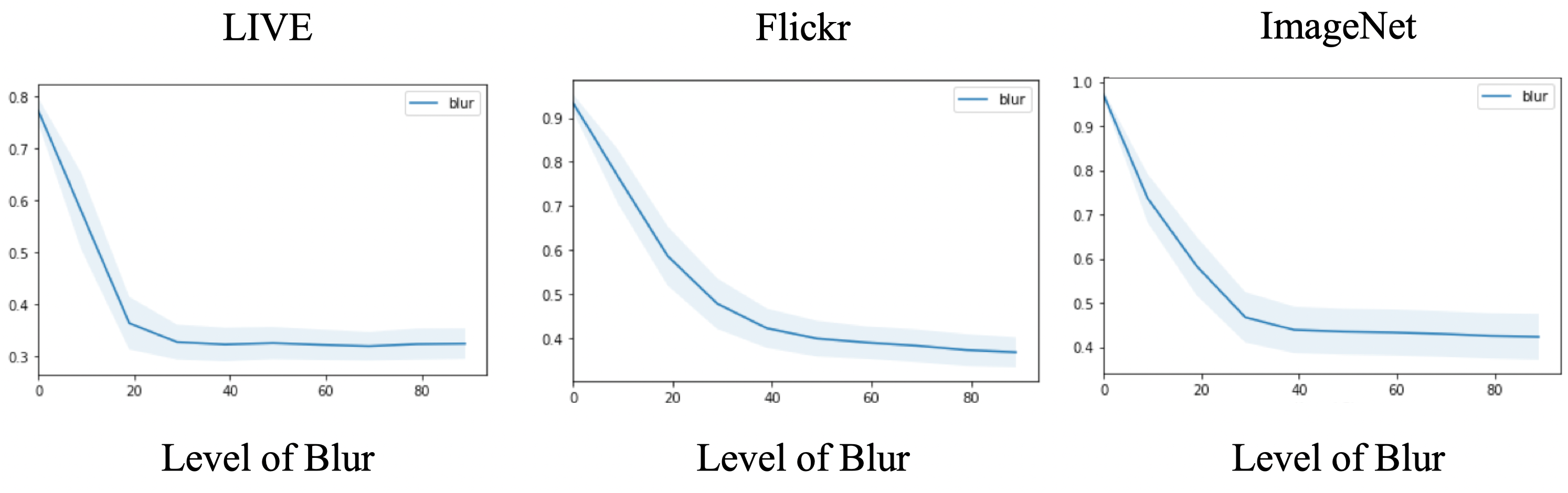}
         \caption{Correlation curves for blurring artifacts}
         \label{fig:12a}
    \end{subfigure}
    \begin{subfigure}[b]{0.5\textwidth}
         \centering
         \includegraphics[width=1.0\textwidth]{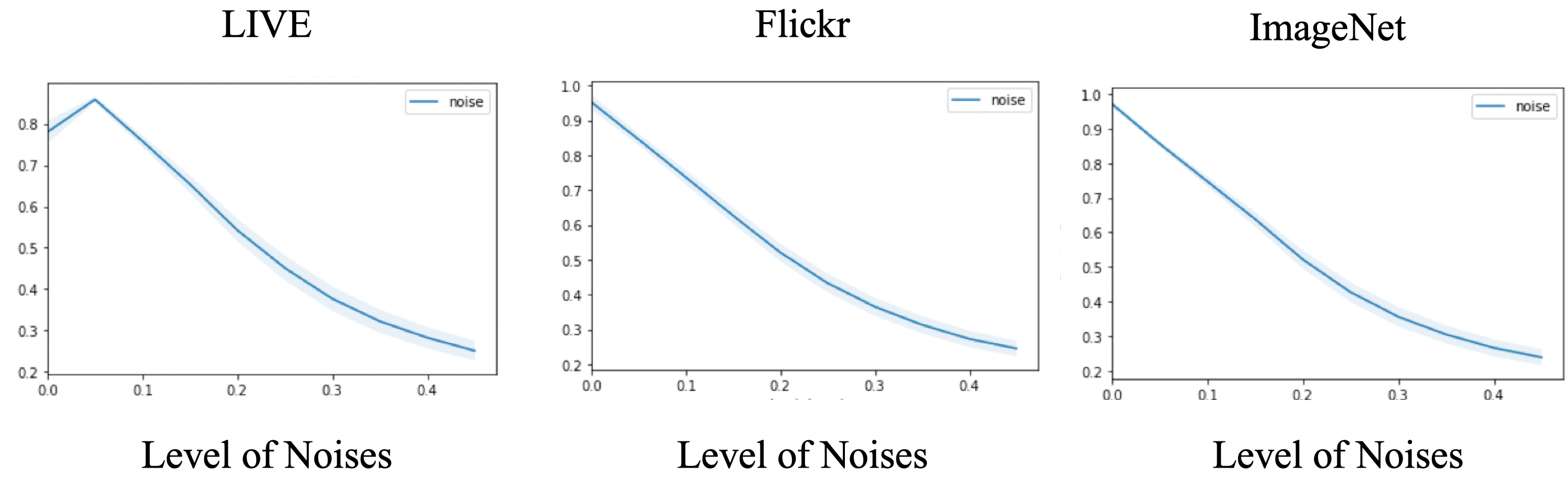}
         \caption{Correlation curves for noise artifacts}
         \label{fig:12b}
    \end{subfigure}
    \caption{Correlation curves between the severity of the corruption artifacts and the QF prediction from the pre-trained network.}
   \label{fig:fig12}
\end{figure*}

As described in section \ref{Relatedwork}, JPEG compression introduces various artifacts such as blurring, blocking, and ringing artifacts. Therefore, we anticipate that our QF Predictor, trained on JPEG-compressed data, can generalize to other types of corruption. To test this hypothesis, we apply the pre-trained RGB QF Predictor to images corrupted by Gaussian Blur and "Salt \& Pepper" noise. The input data preparation process is as follows: we extract a patch from the image, introduce different levels of artifacts to the patch, and feed the corrupted patch to the QF Predictor as illustrated in Figure \ref{fig:fig11}. For each image, we extracted patches from certain randomly-selected locations so that the QF Predictor consistently observes the same region. This approach guarantees a fair comparison, with the only changing factor being the degrees of artifacts rather than the image content. In Figure \ref{fig:fig10}, we present some examples of inference tasks with different levels of artifacts. We expect that the predicted QF will decrease with the corruption.

In Figure \ref{fig:fig12}, we show the correlation curves depicting the relationship between the change in corruption level and the corresponding QF predictions. In most cases, the QF decreases as the level of corruption increases. In each graph, the first half of the curve (where the artifacts are not apparent) shows a steeper decrease with lower variance compared to the latter half. This aligns with our assumption and demonstrates that our network is highly susceptible to subtle artifacts in nearly perfect images, as intended. 

There is one exception observed  in the \textbf{\textit{LIVE}} dataset in \ref{fig:12b}, where the addition of noises seems to slightly boost the QF prediction initially. One possible explanation for this is that original data in \textbf{\textit{LIVE}} naturally have a low image quality. This phenomenon introduces another exciting task: use our model to measure the quality of various current image datasets. 

Table (\ref{tab:t1}) summarizes the application of our QF Predictor to every image in the \textbf{\textit{LIVE}}, \textbf{\textit{Flickr1024}}, and \textbf{\textit{ImageNet}} datasets to estimate the average image quality across each dataset. Our results indicate that the average predicted QF for the \textbf{\textit{LIVE}} data is significantly lower than the other two, suggesting that it is a low-quality image dataset. This finding holds significance considering that images in \textbf{\textit{LIVE}} are already slightly corrupted by various distortion types, including JPEG2000\cite{Authors14}. The validation of our results and the observed differences between datasets reinforce the versatility and reliability of our QF Predictor.

\begin{table}[t]
\centering
\begin{tabular}{||c c c||} 
 \hline
 LIVE & Flickr1024 & ImageNet\\ [0.5ex] 
 \hline\hline
 $0.7800 \pm 0.054$ & $0.9518 \pm 0.043$ & $0.9705 \pm 0.017$ \\ 
 \hline
\end{tabular}
\caption{Predicted average QF for the whole dataset, indicating that images in LIVE generally have a lower quality than Flickr1024 and ImageNet.}
\label{tab:t1}
\end{table}

\begin{figure}[t]
  \centering
   \includegraphics[width=1.0\linewidth]{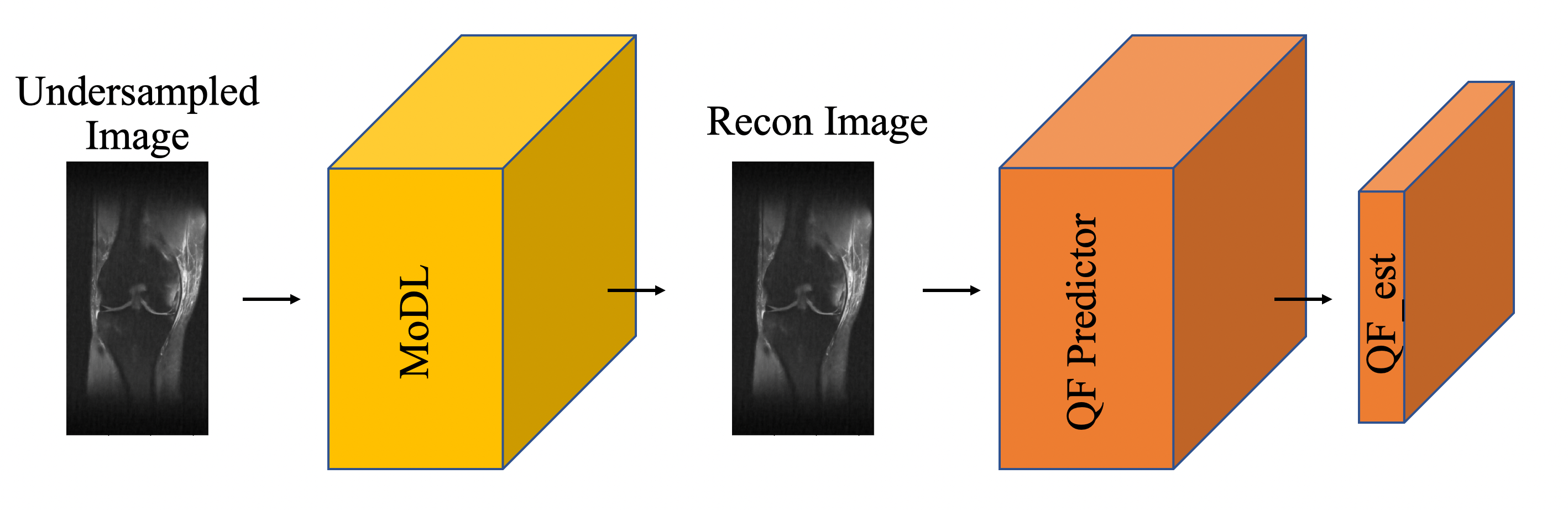}

   \caption{Inference pipeline for grayscale QF Predictor}
   \label{fig:fig13}
\end{figure}

\begin{figure}[t]
  \centering
   \includegraphics[width=1.0\linewidth]{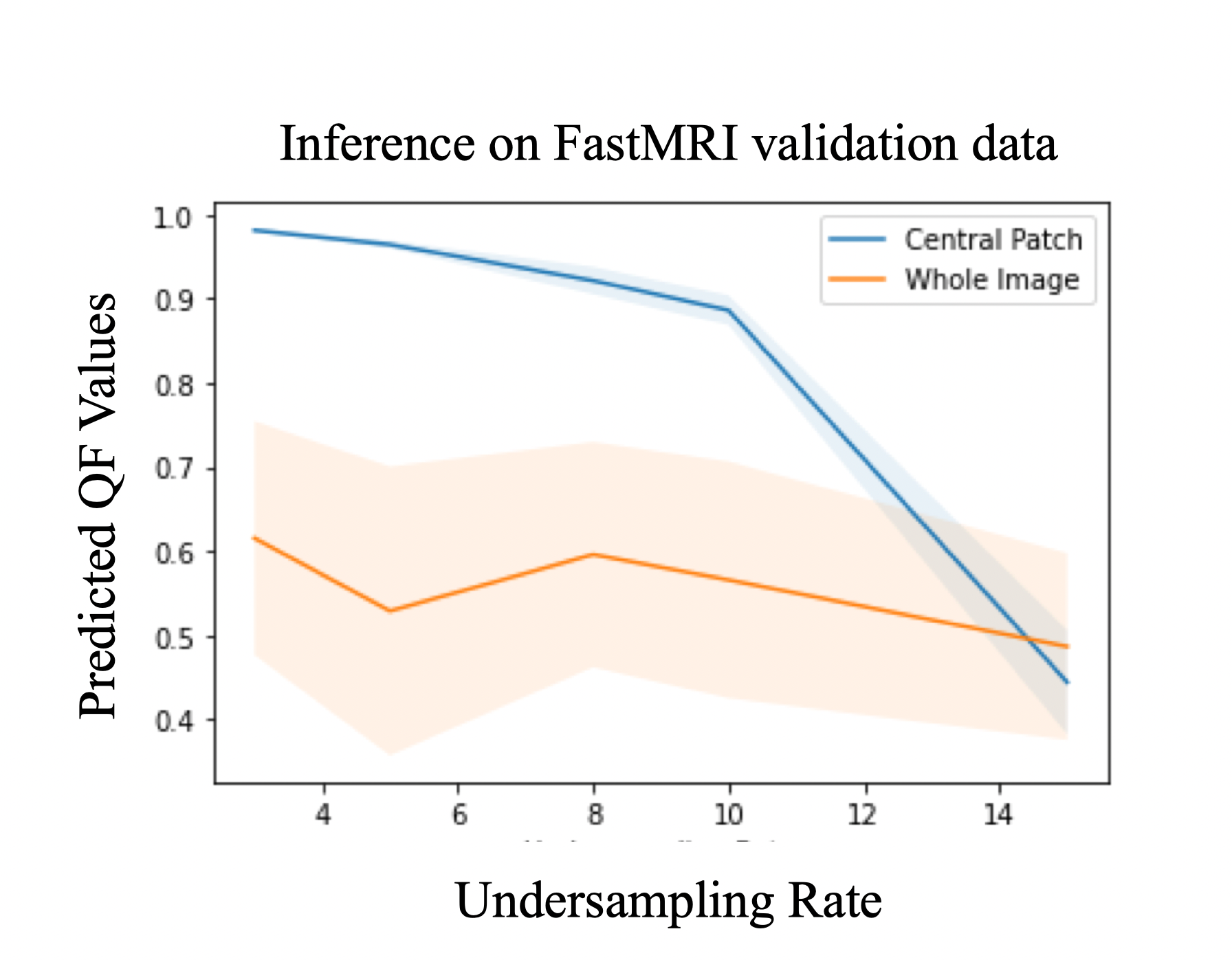}

   \caption{Correlation curves between the under-sampling rate and the QF prediction from the pre-trained network.}
   \label{fig:fig14}
\end{figure}

\begin{figure*}[t]
\centering
   \begin{subfigure}{0.24\linewidth}
    \includegraphics[width=1.0\linewidth]{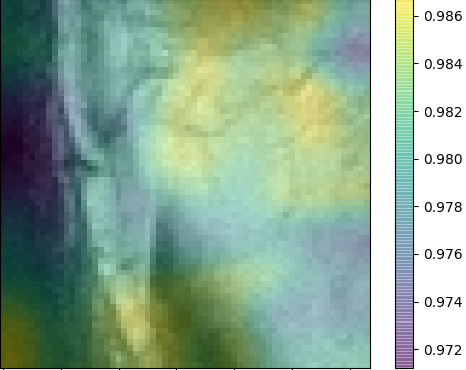}
    \caption{R=3 central patch map} \label{fig:15a}
  \end{subfigure}
  \begin{subfigure}{0.24\linewidth}
    \includegraphics[width=1.0\linewidth]{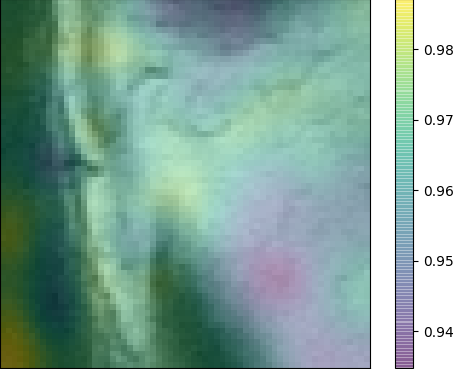}
    \caption{R=5 central patch map} \label{fig:15b}
  \end{subfigure}
  \begin{subfigure}{0.24\linewidth}
    \includegraphics[width=1.0\linewidth]{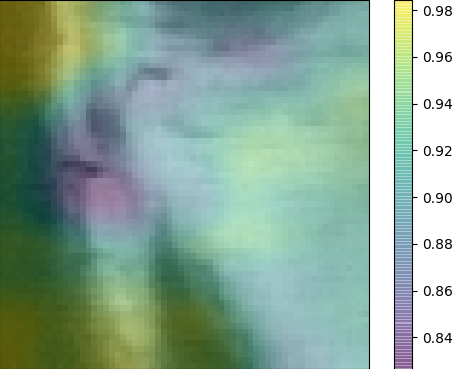}
    \caption{R=8 central patch map} \label{fig:15c}
  \end{subfigure}
  \begin{subfigure}{0.24\linewidth}
    \includegraphics[width=1.0\linewidth]{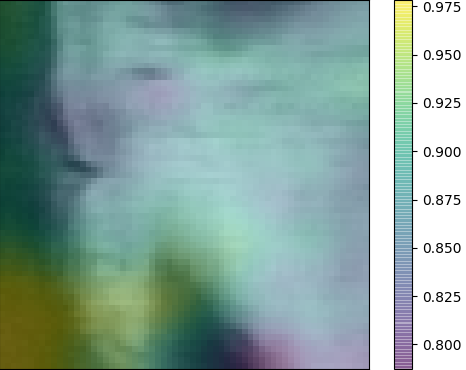}
    \caption{R=10 central patch map} \label{fig:15d}
  \end{subfigure}
  
  \begin{subfigure}{0.24\linewidth}
    \includegraphics[width=1.0\linewidth]{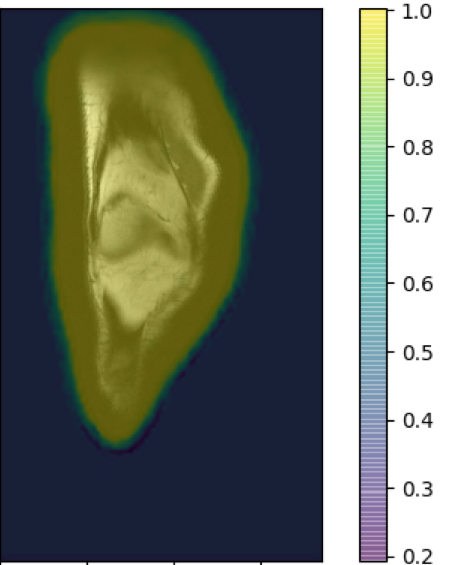}
    \caption{R=3 whole image map} \label{fig:15e}
  \end{subfigure}
  \begin{subfigure}{0.24\linewidth}
    \includegraphics[width=1.0\linewidth]{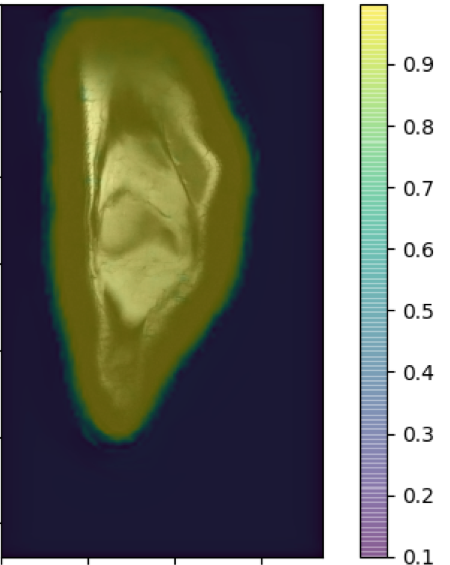}
    \caption{R=5 whole image map} \label{fig:15f}
  \end{subfigure}
  \begin{subfigure}{0.24\linewidth}
    \includegraphics[width=1.0\linewidth]{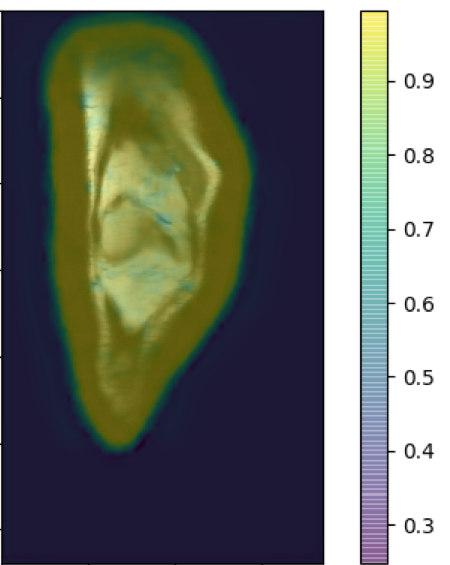}
    \caption{R=8 whole image map} \label{fig:15g}
  \end{subfigure}
  \begin{subfigure}{0.24\linewidth}
    \includegraphics[width=1.0\linewidth]{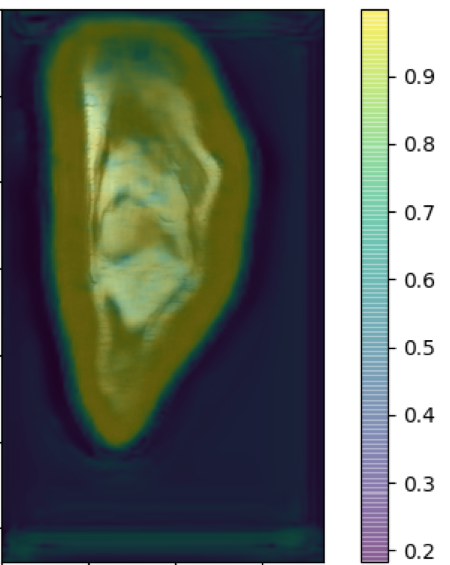}
    \caption{R=10 whole image map} \label{fig:15h}
  \end{subfigure}
  
    \caption{First row: QF predictions mapped to the central patches; Second row: QF predictions mapped to the whole images.}
   \label{fig:fig15}
\end{figure*}

\subsection{Generalization Test to Under-sampling Rate}

Having pre-trained the grayscale QF Predictor on the \textbf{\textit{FastMRI}} knee data (Authors13), our natural curiosity led us to explore whether the network can contribute to improving MRI reconstruction. To achieve this, we first investigate whether the model can predict the under-sampling rate $R$ of the MRI reconstruction. We employ MoDL \cite{Authors11} as the MRI reconstruction network, which generally produces outputs with a low Normalized Root Mean Square Error (NRMSE) and high quality based on SSIM and PSNR metrics. Identifying artifacts in reconstructed MR images can be challenging for individuals without specialized training, except in cases of high under-sampling rates such as $R \geq 10$. However, we anticipate that our QF Predictors, trained with JPEG compression, can generalize to discern subtle imperfections in high-quality MoDL output data.

As shown in Figure \ref{fig:fig13}, we followed a similar pipeline to the RGB inference task. We fed the under-sampled data generated by the Fourier domain mask with varying under-sampling rates into the MoDL network and ran the pre-trained QF Predictor on the high-quality reconstructed data. Note that the output of the MoDL is a complex-value image data, and we need to calculate the corresponding magnitude value to obtain the appropriate image for the QF Predictor. Similar to the RGB generalization test, we expected the QF predictions to decrease as the under-sampling rate increases. However, as shown in Figure \ref{fig:fig14}, although the results were consistent with low variance for the central patch, the performance for whole images exhibited instability with significant uncertainty. 

In Figure \ref{fig:fig15}, we examined the QF Predictor outputs mapped to both the central patch and the whole image. The first row demonstrates that the network can capture subtle corruptions caused by increased under-sampling rates, particularly around the edges of the joint area, in central patches rich with detailed textures. However, in the second row, where we present results for the whole image, the network predominantly focuses on the knee areas and assigns low QF values to near-black background regions. These low values in the 20 to 50 range bias the statistics in the correlation curves displayed in Figure \ref{fig:fig14}, which indicates the limitations of our estimation on the whole image level for MRI reconstruction.

\begin{figure*}[t]
  \centering
   \includegraphics[width=1.05\linewidth]{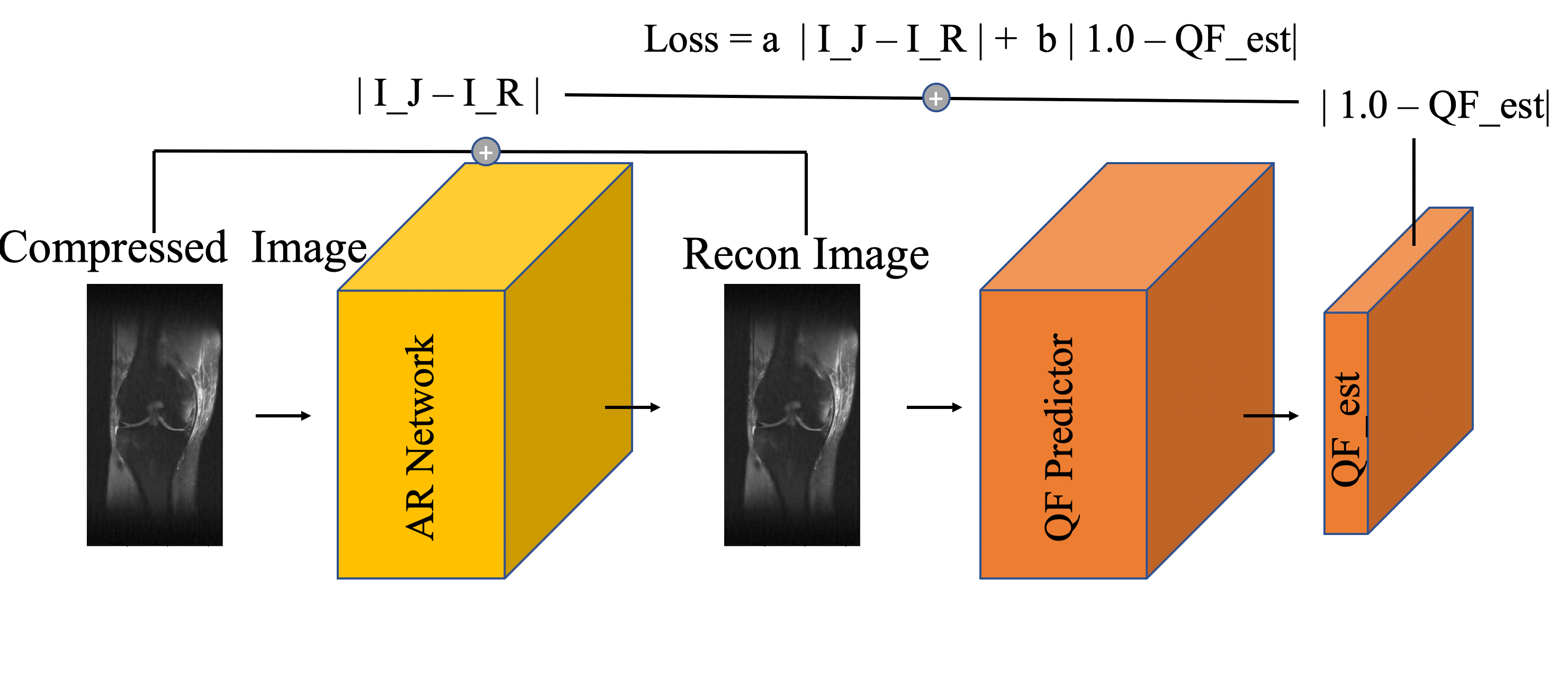}

   \caption{Pipeline for using QF Predictor as a loss function integrated into the image reconstruction networks.}
   \label{fig:fig16}
\end{figure*}

\begin{figure*}[t]
  \centering
  \includegraphics[width=0.8\linewidth]{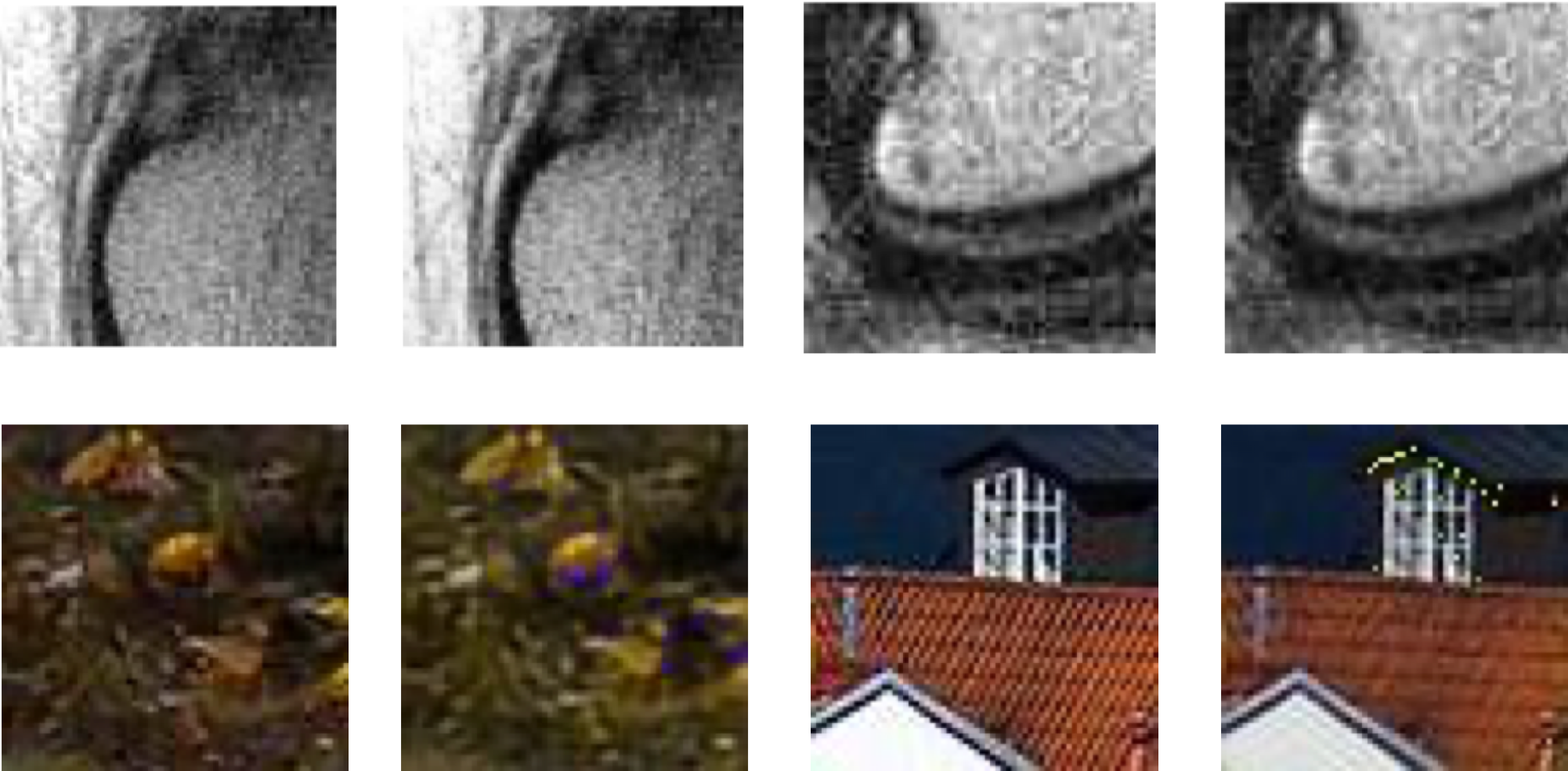}

  \caption{Showcase of input and output pairs, from training with both L1 loss and QF loss term. \textbf{First row}: examples from the grayscale experiments; \textbf{Second row}: examples from the RGB experiments. }
  \label{fig:fig17}
\end{figure*}

\subsection{Application as a Perceptual Loss Function}
In recent Computer Vision research, researchers have often incorporated perceptual loss terms, such as SSIM, into their models to generate more visually pleasing results. As shown previously, our QF Predictor can capture subtle differences between perfect-quality and high-quality images that are sometimes challenging for humans to perceive. This observation has inspired us to utilize the Quality Factor network as a deep perceptual loss function to train an image reconstruction network in cases where  ground truth data is unavailable. For our artifact removal task, we used the SR-CNN\cite{Authors15} and AR-CNN\cite{Authors16} as the backbone networks. As this task can be applied to both RGB and grayscale images, we employed pre-trained grayscale and RGB QF predictors to estimate the image quality of the reconstruction output for different types of images. As shown in Figure \ref{fig:fig16}, we incorporated both a data consistency loss term between the input and the output of the image reconstruction network and a perceptual term based on quality factor estimations. 

Figure \ref{fig:fig17} shows several examples from both RGB and grayscale experiments. The graph illustrates that the QF loss in the grayscale artifacts removal task resulted in some minor improvements, such as sharpening edges and reducing general noise levels, but it did not lead to noticeable visual quality enhancement. However, in the corresponding artifacts removal task on RGB images, the QF loss term introduced some additional inconsistencies, such as the changes in color tones and the emergence of generative artifacts, such as the purple dots in the first RGB pair and yellow dots in the second RGB pair. This suggests that the QF Predictor somehow overrides the image reconstructor's prior knowledge and focuses solely on image quality without considering data consistency. Achieving a balance point between the data consistency loss term and the QF term becomes crucial, requiring meticulous and rigorous weightings to prevent erosion of reconstruction quality.

\section{Conclusions}
This work presents a self-supervision method based on JPEG compression for a reference-free deep quality metric prediction model. Our proposed model does not rely on extensive training datasets, yet it is capable of capturing perceived image quality similar to humans and distinguishing small nuances between near-perfect and lossless images. We have demonstrated that a network with only seven convolution layers can be trained and achieve convergence within a short training period, providing accurate predictions with low variance on RGB and grayscale validation data. Furthermore, we have showcased the model's ability to generalize to other tasks, such as predicting corruption levels induced by Gaussian blur and "Salt \& Pepper" noise, as well as estimating the under-sampling rate of high-quality MRI reconstruction network outputs for detailed regions. 

However, when applying the model as a loss term to visually improve the output of image artifact removal networks, we observed sub-optimal performance. This is attributed to the tradeoff between an increase in image quality, and the potential risk of compromising data consistency and introducing generative artifacts. Therefore, careful consideration and rigorous weighting between the data consistency loss term and the quality factor term are necessary to strike a balance that ensures both image quality improvement and preservation of reconstruction accuracy.

\section{Acknowledgements}
This project is under the guidance of Alfredo De Goyeneche, Efrat Shimron, and Michael (Miki) Lustig. We also thank Ke Wang and Stella X. Yu for their insightful advice.

\newpage
{\small
\bibliographystyle{unsrt}
\bibliography{egbib}
}
\end{document}